\documentclass[twocolumn,showpacs,preprintnumbers,amsmath,amssymb]{revtex4}
\usepackage{graphicx}
\input epsf \unitlength=1mm

\begin{document}

\title{Extended Hartree-Fock method based on pair density functional theory}

\author{Bal{\'a}zs Het{\'e}nyi$^{1,2}$\footnote{Present address:Institut
    f\"ur Theoretische Physik, Technische Universit\"at Graz, Petersgasse 16,
  A-8010, Graz, Austria} and Andreas W. Hauser$^3$}

\affiliation{$^1$Institut f\"ur Theoretische Physik, Technische Universit\"at Graz, Petersgasse 16, A-8010, Graz, Austria \\
  $^2$SISSA-International School of Advanced Studies, via Beirut 2-4,
  Trieste,
  I-34014, Italy \\
  $^3$Institut f\"ur Experimentalphysik, Technische Universit\"at Graz,
  Petersgasse 16, A-8010, Graz, Austria }

\begin{abstract}
  A practical electronic structure method in which a two-body functional is
  the fundamental variable is constructed.  The basic formalism of our method
  is equivalent to Hartree-Fock density matrix functional theory [M. Levy in
  {\it Density Matrices and Density Functionals}, Ed. R. Erdahl and V.~H.
  Smith Jr., D. Reidel, (1987)].  The implementation of the method consists
  of solving Hartree-Fock equations and using the resulting orbitals to
  calculate two-body corrections to account for correlation.  The correction
  terms are constructed so that the energy of the system in the absence of
  external potentials can be made to correspond to approximate expressions
  for the energy of the homogeneous electron gas.  In this work the
  approximate expressions we use are based on the high-density limit of the
  homogeneous electron gas.  Self-interaction is excluded from the two-body
  functional itself.  It is shown that our pair density based functional does
  not suffer from the divergence present in many density functionals when
  homogeneous scaling is applied.  Calculations based on our pair density
  functional lead to quantitative results for the correlation energies of
  atomic test cases.
\end{abstract}

\maketitle
\section{Introduction}

\label{sec:intro}

The most commonly used methods in electronic structure for atoms and
molecules are density-functional
theory~\cite{Hohenberg64,Kohn65,Parr89,Dreizler90} (DFT) and
Hartree-Fock~\cite{Szabo96,Young01} (HF) theory.  The former is based on the
Hohenberg-Kohn theorems, which state that all ground state observables, in
particular the energy can be written as a functional of the one-body
density~\cite{Hohenberg64}.  Since the functional itself is unknown, in
actual implementations the system of interacting electrons is mapped onto an
auxiliary system of non-interacting electrons resulting in the Kohn-Sham
single-particle equations~\cite{Kohn65}.  This ansatz implicitly assumes that
all densities originating from anti-symmetric many-body wave-functions can be
represented by a non-interacting wave-function.  The Kohn-Sham energy
functional includes a correction term, known as the exchange-correlation
energy, which accounts for effects of exchange and correlation, and is
usually approximated using the theory of the homogeneous electron
gas~\cite{Dreizler90,Senatore94,Ceperley80,Mahan00,Mattuck67,March67,GellMann57,Nozieres58}
(HEG).  This correction term usually depends on the one-body density,
although orbital-dependent variations also exist~\cite{Perdew81}.

The HF approximation~\cite{Szabo96,Young01} is based on the use of a Slater
determinant for the many-body wave-function, hence the work-horse equations
are also single-particle equations (as in DFT), but exchange is incorporated
in this case. Correlation effects are neglected.  There are a variety of
methods which incorporate correlation effects into HF, such as use of a
linear combination of Slater determinants~\cite{Young01}, perturbation
theory~\cite{Szabo96,Young01}, or a combination thereof, but all at the cost
of greatly increasing computational demand~\cite{Foresman96} (both methods
involve the use of virtual orbitals, increasing the size of the Hilbert space
in which the wave-function is calculated).  The connections between HF and
DFT have also been investigated~\cite{Payne79,Levy79} and it has been shown
that the Hohenberg-Kohn theorems apply within the Hilbert space of Slater
determinants, hence it is in principle possible to construct a DFT that
corresponds to HF.

There has also been active interest in extending DFT by using the
pair density~\cite{Ziesche96,Levy01,Furche04,Hetenyi04,Nagy02,Nagy04,Nagy06,
  Percus05,Ayers05,Ayers06a,Ayers06b,Ayers07,Gonis96,Gonis97,Pistol04,Higuchi07} (pair density functional theory (PDFT))
or the density matrix~\cite{Zumbach85,Levy87,Mazziotti06,Coleman00,Coleman63}
as the fundamental variable. Most PDFT studies have dealt with formal issues,
an example of an exception being the work of Gonis {\it et
  al.}~\cite{Gonis96} where an implementation is developed in the context of
strongly correlated systems.  The development of a practical electronic
structure method based on PDFT, and applicable to atomic and molecular
systems in general appears to be lacking.

PDFT is an extension of the standard DFT in that the Hohenberg-Kohn
theorems~\cite{Hohenberg64} are generalized to the two-body (or $N$-body)
density~\cite{Ziesche96,Levy01,Furche04,Hetenyi04,Nagy04,Percus05,Ayers05,Ayers07,Gonis96}.
Application of the Kohn-Sham ansatz~\cite{Kohn65} is not straightforward in
the case of PDFT, since a non-interacting pair-density, such as that obtained
from a Slater determinant can not represent all possible pair densities
arising from an arbitrary anti-symmetric wave-function.  The lack of
correlation between electrons with anti-parallel spin presents an obstacle to
representing an arbitrary anti-symmetric many-body wave-function via Slater
determinants.

An approach related to PDFT based on reduced density matrices (RDM) is also
under considerable
investigation~\cite{Zumbach85,Levy87,Mazziotti06,Coleman00,Coleman63}.  Levy
has demonstrated~\cite{Levy87} that a correction term due to correlation is a
unique functional of the Hartree-Fock density matrix.  Levy has also
suggested~\cite{Levy87} possible starting points for the construction of such
correlation corrections.

In this work we construct a PDFT-based method that is robust, simple, and it
can be implemented in an HF context.  We argue that an {\it exact} energy
functional can in principle be constructed based on the HF pair density.
Since the HF pair density and one-particle RDM are related one-to-one, our
formalism is equivalent to that derived by Levy.~\cite{Levy87}  A pair density
based correction to HF has also recently been suggested by Higuchi and
Higuchi~\cite{Higuchi07}.

We also construct approximations to the exact correlation functional via
generalizing known scaling relations~\cite{Levy01}, which allow the
construction of pair density dependent correlation functionals based on the
theory of the homogeneous electron gas
(HEG)~\cite{Dreizler90,Senatore94,Ceperley80,Mahan00,Mattuck67,March67,GellMann57,Nozieres58}.
The method suggested by Higuchi and Higuchi~\cite{Higuchi07} uses the
standard scaling relations derived by Levy and Ziesche~\cite{Levy01}.  In
particular, in this work, we construct correlation functionals which
correspond to expressions valid for the HEG in the high-density
limit~\cite{Mahan00,Mattuck67,March67,GellMann57,Nozieres58} (expected to be
a good approximation for the atomic test cases presented herein).  Our
procedure can, in principle, be extended to other functional forms (power
series in the density, such as the low-density form) for the correlation
energy es well.  Since our method amounts to implementing a correction term
to an HF calculation (to account for correlation), and it does not require
the use of virtual orbitals, the scaling with system size is that of HF
itself.  We apply our method to calculate atomic energies and obtain
quantitative agreement with known correlation energies.

We also investigate the properties of our constructed functionals under
homogeneous coordinate scaling.  DFT correlation energy functionals with
logarithmic terms~\cite{Mahan00,Mattuck67,March67,GellMann57,Nozieres58} are
known to be divergent when the scaling parameter is made
infinite~\cite{Perdew81b,Gorling93}.  This behavior contradicts the
known~\cite{Perdew81b,Levy89,Gorling92} scaling behavior of the actual
correlation energy which is bounded below.  It is shown that our correlation
functionals do not exhibit this deficiency.

This paper is organized as follows.  In Section \ref{sec:formal} the formal
development is presented.  We describe a procedure to construct correlation
functionals based on known scaling relations in PDFT, and comment on the
properties of the kinetic energy functional.  The explicit construction of
energy functionals is explained in section \ref{sec:functionals}.  In section
\ref{sec:calculation} we discuss the details of our calculations.  Section
\ref{sec:results} consists of our results and analysis.  We conclude our work
in section \ref{sec:conclusion}.

\section{Formal development}

\label{sec:formal}

\subsection{Two-body density-functionals}

\label{ssec:representability}

The ground state energy of a many-body fermionic system has been shown to be
a unique functional of the single-particle many-body density
matrix~\cite{Zumbach85,Dreizler90}.  In Ref.  \onlinecite{Levy87} it was
shown that there exists an additive correction to the HF energy functional
that depends on the HF one-body RDM.  The proof is based on investigating the
correspondence between the HF one-body RDM and the one-body potential of a
given system.  In particular an exact formula is provided by which one can
determine the external potential as a function of the HF orbitals.

In standard DFT, an applicable method is constructed via the Kohn-Sham
ansatz~\cite{Kohn65}.  This ansatz involves a mapping between the interacting
system of electrons to an auxiliary non-interacting one, enabling the use of
single-particle equations.  The solutions of the single-particle equations
are single-particle orbitals, and one must choose a way of writing the
density in terms of these single particle orbitals.  The most often used
choice is the one that satisfies the Pauli principle
\begin{equation}
  \rho({\bf r}) = \sum_{i,\sigma} f_{i,\sigma}|\phi_{i,\sigma}({\bf r})|^2,
\label{eqn:rho_def}
\end{equation}
where $\rho({\bf r})$ denotes the density $f_{i,\sigma}$ denotes an
occupation number and $\phi_{i,\sigma}({\bf r})$ denotes a particular
spin-orbital.  Usually $f_{i,\sigma}=1$ which corresponds to the unrestricted
HF density.  To our knowledge it is not proved that the non-interacting form
for the density based on a single determinant can represent all interacting
densities corresponding to anti-symmetric many-body wave-functions.

While standard DFT was made into an applicable electronic structure method by
the Kohn-Sham ansatz, the generalization of the ansatz to the case of
two-body functionals such as the pair density is not straightforward, due to
representability issues~\cite{Ayers05,Ayers06a,Ayers06b,Ayers07,Pistol04}.
Writing a {\it non-interacting} energy-functional requires the definition of
the pair density $n({\bf r},{\bf r'})$ in terms of the orbitals of the
non-interacting system.  An obvious initial choice is
\begin{eqnarray}
\nonumber
n_{HF}({\bf r},{\bf r'}) &=& 
\sum_{i,j,\sigma,\sigma'}\{|\phi_{i,\sigma}({\bf r})|^2
        |\phi_{j,\sigma'}({\bf r'})|^2  \\
 &-& \phi_{i,\sigma}({\bf r})\phi_{i,\sigma}({\bf r'})
  \phi_{j,\sigma'}({\bf r})\phi_{j,\sigma'}({\bf r'})
\}
\label{eqn:n_def}
\end{eqnarray}
obtained from a Slater determinant (also known as the Hartree-Fock
pair density).  When this definition for the pair density is used, it can be
shown that the motion of electrons with anti-parallel spins is
uncorrelated~\cite{Szabo96}.  Since this is not so for the anti-symmetric
many-body wave-function, exact representation of an interacting system by a
non-interacting one does not appear to be possible.  In the theory of
correlated wave-functions this effect is usually implemented via the cusp
condition for correlation factors~\cite{Ceperley80}.

\begin{center}
\begin{table}
\begin{tabular}{|l|l|l|l|}
\hline
             & Gell-Mann and Brueckner  & Nozi\`eres and Pines  & RPA \\ \hline
 $A_c$         & $-0.048$               & $-0.058$     & $-0.071$ \\ \hline
 $B_c$         & $+0.0311$               & $+0.016$     & $+0.0311$ \\ \hline
 $A$         & $-0.0619$               & $-0.0649$     & $-0.0859$ \\ \hline
 $B$         & $-0.0104$               & $-0.00517$     & $-0.0104$ \\ \hline
 $\tilde{A}$ & $-0.0598$               & $-0.0639$     & $-0.0838$ \\ \hline
 $\tilde{B}$ & $-0.0622$               & $-0.031$     & $-0.0622$ \\ \hline
\end{tabular}
\caption{Table of constants for different correlation energy functionals (Har)
  (functional forms are given in Eqs. (\ref{eqn:ec_dft}) and 
  (\ref{eqn:ec_pdft})).}
\label{tab:constants}
\end{table}
\end{center}

On the other hand, using the result of Levy~\cite{Levy87} one can easily
argue that the energy is a unique functional of the Hartree-Fock pair density
since the density matrix and the Hartree-Fock pair density are in a
one-to-one relation.  Hence the correlation energy may be written
\begin{equation}
E_c[n_{HF}] = \tilde{E}[n_{HF}] - E_{HF}[n_{HF}],
\label{eqn:exact_Ec}
\end{equation}
i.e. as a unique functional of the HF two-body density (in Eq.
(\ref{eqn:exact_Ec}) $\tilde{E}[n_{HF}]$($E_{HF}[n_{HF}]$) denote the exact
energy(Hartree-Fock energy) as a functional of the Hartree-Fock pair density
$n_{[HF]}$).  Below we construct approximations for $E_c[n_{HF}]$.

Here, as in the case of DFT, arguments based on mappings establish one to one
relations between observables (ground state energy) and coordinate-dependent
functionals (density), however the exact functional relations are not given
by the formalism.  While in the Kohn-Sham DFT method a mapping between a
non-interacting and interacting system is invoked, where the two are
constrained to have the same density, in the formalism developed above one
does not have the two-body density readily available.  Although, in
principle, a perturbative approach carried out to infinite order gives an
exact solution, in practice one has to content oneself with an approximate
solution to this problem.  Higuchi and Higuchi~\cite{Higuchi07} suggest
minimizing the energy of an augmented Hartree-Fock equation, but this method
can only be approximate due to lack of opposite-spin correlation.  Possible
alternative approaches are perturbation
theory~\cite{March67,Mattuck67,Kimball76}, two-body generalization of
plane-wave perturbation theory~\cite{March67}, or variational approaches such
as the Fermi hyper-netted-chain~\cite{Krotscheck71,Fantoni74} method.  A
two-body density can also be obtained from an HF solution via constructing a
Jastrow-Slater wave-function to account generally for correlation, and the
relevant observables could be evaluated by quantum Monte Carlo
methods~\cite{Foulkes01}.  In all of these methods, the HF wave-function
serves as input, and the output consists of the many-body properties.

Unfortunately our formalism does not provide an inversion formula that can be
used to determine the two-body potential from the two-body density in the HF
case.  This is due to the fact that, although the HF two-body density is a
two-body object, it is composed of single-particle orbitals.
If a particular two-body potential is assumed (Coulomb repulsion
between electrons) then the inversion formula presented by
Levy~\cite{Levy87} can be used to obtain the one-body potential.

The method constructed from the above reasoning requires an HF calculation,
supplemented by a correction term due to correlation.  The correction term is
a functional of the HF pair density.  Knowledge of the HF wave-function gives
exact information about the ground state energy (and by extension all ground
state observables) in principle, and it is not necessary to calculate the
exact wave-function, if one is interested only in ground state expectation
values.  This can not be done exactly, due to the lack of the exact
functional form, and approximation schemes need to be conceived for the
particular set of observables in mind.  In this study we develop such a
scheme for the ground state energy, or more specifically the correlation
energy.

\subsection{Generalized scaling relations and sum rules}

\label{ssec:scaling}

To construct correlation functionals $E_c[n_{HF}]$ we write the scaling
relations due to Levy and Ziesche~\cite{Levy01}.  In Ref.
\onlinecite{Levy01} it was shown that possible terms that contribute to an
expansion of the kinetic energy functional (per particle) of the form
\begin{equation}
  t[n] = \frac{1}{N}\sum_b A_b \int d {\bf r} d {\bf r'}  n({\bf r},{\bf
    r'})^a |{\bf r} - {\bf r'}|^b,
 \label{eqn:T_ser}
\end{equation}
must be such that the relation $6a-b=8$ is satisfied.  To derive the
HF kinetic energy functional for the {\it homogeneous} case we
substitute the relation
\begin{equation}
  n_{HF}({\bf r},{\bf r'}) = \rho^2 g(|{\bf r} - {\bf r'}|),
 \label{eqn:n_g}
\end{equation}
into Eq. (\ref{eqn:T_ser}).  In Eq. (\ref{eqn:n_g}) $\rho$ denotes the
one-body density, and $g(r)$ denotes the radial distribution function.  Using
the dimensionless distance scaled by the Fermi wave-vector $x=k_F r$, where
$k_F = (3 \pi^2 \rho)^{\frac{1}{3}}$, elementary manipulations lead to the
kinetic energy expression 
\begin{equation}
  t[n_{HF}] = \rho^{\frac{2}{3}} \sum_b \tilde{A_b} \int_{0}^{\infty} d x x^{b+2}
  g(x)^a,
\label{eqn:t_g}
\end{equation}
where~\cite{Mahan00,Dreizler90}
\begin{equation}
 g(x) = \{
1 - \frac{9}{2 x^6}(\mbox{sin}(x)-x\mbox{cos}(x))^2 
\}.
\label{eqn:gxu}
\end{equation}
The $\tilde{}$ on $\tilde{A_b}$ indicates that some constants have been
absorbed into $\tilde{A_b}$.  It is
well-known~\cite{Mahan00,Dreizler90,Parr89} that the kinetic energy in the HF
approximation can be written as
\begin{equation}
  t[n_{HF}] = \frac{3}{10}(3 \pi^2)^{\frac{2}{3}} \rho^{\frac{2}{3}},
\label{eqn:t_hf}
\end{equation}
thus, from Eqs. (\ref{eqn:t_g}) and (\ref{eqn:t_hf}) a sum rule is
established on the coefficients $\tilde{A_b}$ for the homogeneous
non-interacting case, which we write as
\begin{equation}
 \sum_b \tilde{A_b} \int_{0}^{\infty} d x x^{b+2}
  g(x)^a  = \frac{3}{10}(3 \pi^2)^{\frac{2}{3}}.
\label{eqn:sum_rule1}
\end{equation}
Regarding an homogeneous interacting system we can also conclude from the
above analysis that if the kinetic energy exhibits scaling other than
$\rho^{\frac{2}{3}}$ then the scaled radial distribution function $g(x)$ also
depends on the density.

It is possible and useful to generalize the scaling relations of Levy and
Ziesche~\cite{Levy01} so that a particular functional can reproduce a given
density dependence when a homogeneous non-interacting system is invoked.  A
general functional of the pair density $n$ may be written
\begin{equation}
  f(\Gamma)[n] = \frac{1}{N}\sum_{a} C_a
\int d{\bf r} d{\bf r'} n({\bf r},{\bf r'})^a |{\bf r} - {\bf r'}|^b,
\label{eqn:f}
\end{equation}
where the condition $6a-b=6+\Gamma$ has to be satisfied.  For the homogeneous
non-interacting case (i.e. when Eq. (\ref{eqn:n_g}) is substituted for
$n_{HF}$) $f(\Gamma)[n]$ will exhibit a density scaling
\begin{equation}
  f(\Gamma)[n_{HF}] = \tilde{C}_{\Gamma} \rho^{\Gamma},
\label{eqn:f_HF}
\end{equation}
resulting in the generalized sum rule
\begin{equation}
  \tilde{C}_{\Gamma} = \sum_a \tilde{C}_a \int_{0}^{\infty} d x x^{b+2} g(x)^a,
\label{eqn:sum_rule2}
\end{equation}
valid for the auxiliary non-interacting system.  The $\tilde{}$ on
$\tilde{C}(a)$ indicates that some constants have been absorbed into
$\tilde{C}(a)$.  Using Eq. (\ref{eqn:f_HF}) one can construct pair density
analogs for correlation functionals that are given in terms of power series
in the density $\rho$ (examples are the low and high-density approximations).
It is also possible to obtain sum rules of the form Eq. (\ref{eqn:sum_rule2})
valid for the correlation energy.  

\subsection{Properties of the kinetic energy functional}

\label{ssec:kinetic}

The properties of the integral in Eq. (\ref{eqn:t_g}) can now be studied since
the functional form of the radial distribution function is known (Eq.
(\ref{eqn:gxu})).  The behavior of $g(x)$ at zero and infinity place bounds on
the exponent $a$ (and hence $b$).  At large distances $g(x)^a$ approaches
$1$, hence 
\begin{equation}
b+2<-1.  
\label{eqn:a1}
\end{equation}
Close to zero, however $g(x)$ for the unpolarized case approaches a constant
($\frac{1}{2}$) which leads to the relation
\begin{equation}
b+2>-1,  
\label{eqn:a2}
\end{equation}
which is inconsistent with Eq. (\ref{eqn:a1}). Hence for the spin-unpolarized
case the integral is divergent.  This divergence is due to the radial
distribution function being finite at the origin, in the case of the { \it
  ideal} Fermi gas.  

The divergence can be eliminated in several ways.  One is to assume that the
kinetic energy is a sum of terms, one for each electron spin component.  The
radial distribution function corresponding to one spin component only is zero
at the origin, therefore the divergence does not arise.  In HF this procedure
is implicit.  Corrections to the energy functional can also be constructed in
terms of the exchange hole, as is done below.

\section{Construction of energy functionals}

\label{sec:functionals}

Since the HF method is now well-established, our principal aim is to
construct energy functionals that are tractable by equivalent numerical
methods.  To achieve this aim, we construct approximations for the
correlation energy $E_c$ in terms of the Hartree-Fock pair density.  Since in
the HF theory of the HEG the quantity that appears is the so called
exchange-correlation hole defined as
\begin{equation} 
  n_{xc}({\bf r},{\bf r'}) = n({\bf r},{\bf r'}) - \rho({\bf r})\rho({\bf r'}),
\label{eqn:xc_hole}
\end{equation} 
we find it more convenient to use $n_{xc}$ as our fundamental quantity (input
function).  Below we derive approximations from the theory of the HEG, and
in HEG the energy would diverge without the constant positive background.
The second term in Eq. (\ref{eqn:xc_hole}) corresponds to the positive
background, hence the divergence is canceled.  The scaling relations derived
in the previous section are valid for $n_{xc}$ as well, since both terms on
the right-hand side of Eq. (\ref{eqn:xc_hole}) exhibit the same coordinate
scaling.  When the HF approximation is invoked the exchange correlation hole
becomes the exchange hole
\begin{equation}
n_{x}({\bf r},{\bf r'}) =
 - \sum_{i,j,\sigma} \phi_{i,\sigma}({\bf r})\phi_{i,\sigma}({\bf r'})
  \phi_{j,\sigma}({\bf r})\phi_{j,\sigma}({\bf r'}).
\label{eqn:x_hole}
\end{equation}
Eq. (\ref{eqn:x_hole}) reintroduces self-interaction which can be excluded
here by restricting the summation to $i\neq j$, 
\begin{equation}
\tilde{n}_{x}({\bf r},{\bf r'}) =
 - \sum_{i \neq j,\sigma} \phi_{i,\sigma}({\bf r})\phi_{i,\sigma}({\bf r'})
  \phi_{j,\sigma}({\bf r})\phi_{j,\sigma}({\bf r'}),
\label{eqn:x_hole2}
\end{equation}
where the prime indicates the self-interaction corrected (SIC) sum.  In the
two-body density functional theory presented here self-interaction is thus
excluded from the density functional itself.  In DFT SIC is usually achieved
by subtracting energies (LSD energy functionals for Hartree, exchange and
correlation) corresponding to single-particle densities~\cite{Perdew81}.
Recently, Lundin and Eriksson~\cite{Lundin01} have proposed a SIC procedure
where the self-interaction is addressed by subtracting single-particle
densities from the total density.  

It has been pointed out based on scaling arguments in the case of the
exchange energy that self-interaction terms disappear in the thermodynamic
limit~\cite{Bobel83,Rae75}.  The reason for the disappearance of the
self-interaction terms is that self-interaction terms have to scale as $N$
with the number of particles, whereas the overall number of interactions
between different particles scales as $N(N-1)/2$.  An underlying assumption
of the argument which is valid for the exchange energy is a single-particle
picture in which one can decompose the interactions between different pairs
of particles.  Since the correlation energy in the high-density limit which
serves as the basis for our approximations here is also a single-particle
based expression (the diagrammatic summation gives an operator whose
expectation value has to be evaluated over a Slater
determinant~\cite{GellMann57,Nozieres58,Mahan00}), and it is also valid in
the thermodynamic limit the arguments in Refs.  ~\onlinecite{Bobel83,Rae75}
are valid for the correlation functionals used here (below Eq.
(\ref{eqn:e_c_dft})) as well.  Thus the expressions used below for the
correlation energy (Eq. (\ref{eqn:e_c_dft})), valid in the thermodynamic
limit, do not exhibit self-interaction.  It can also be expected that use of
the SIC exchange hole to reproduce the functionals such as Eq.
(\ref{eqn:e_c_dft}) is thus advantageous, since it lacks self-interaction by
construction.

\begin{center}
\begin{table}
\begin{tabular}{|l|l|l|l|l|l|}
  \hline
  $E_c$      & He     &   Be   & Ne    &  Mg   & Ar    \\ \hline
  $RPA $     & 0.130  & 0.251  & 0.884 & 1.17  & 1.69  \\ \hline
  $GB $      & 0.0821 & 0.154  & 0.644 & 0.782 & 1.26  \\ \hline
  $NP $      & 0.109  & 0.213  & 0.662 & 0.935 & 1.24  \\ \hline
  $RPA-SIC-LDA$ & 0.014  & 0.033  & 0.198 & 0.222  & 0.398  \\ \hline
  $GB-SIC-LDA $ & 0.014 & 0.033  & 0.198 & 0.222 & 0.398  \\ \hline
  $NP-SIC-LDA $ & 0.0072  & 0.016  & 0.0987 & 0.111 & 0.198   \\ \hline
  $RPA-SIC-LSD$ & 0.082  & 0.161  & 0.589 & 0.689  & 1.13  \\ \hline
  $GB-SIC-LSD $ & 0.058 & 0.113  & 0.469 & 0.545 & 0.912  \\ \hline
  $NP-SIC-LSD $ & 0.063  & 0.124  & 0.404 & 0.476 & 0.761  \\ \hline
  $VMC$~\cite{Buendia06}  & - & 0.073  & 0.346 & 0.372 & 0.576 \\ \hline
  Clementi and  &0.045 & 0.094  & 0.3870& 0.438 & 0.722 \\ 
  Hoffmann~\cite{Clementi95}  & &   & &  & \\ \hline 
\end{tabular}
\caption{Correlation energies (Har) from DFT calculations using various 
  correlation functionals and their self-interaction corrected 
  versions for closed-shell atoms.}
\label{tab:DFT}
\end{table}
\end{center}

\begin{center}
\begin{table}
\begin{tabular}{|l|l|l|l|l|l|}
  \hline
  $E_c$      & Ca    &   Zn   & Kr    &  Xe     \\ \hline
  $RPA $     & 1.85  & 3.15   & 3.88  & 6.12    \\ \hline
  $GB $      & 1.37  & 2.43   & 3.01  & 4.83    \\ \hline
  $NP $      & 1.36  & 2.23   & 2.73  & 4.24    \\ \hline
  $RPA-SIC-LDA$ & 0.429  & 0.822  & 1.01  & 1.66    \\ \hline
  $GB-SIC-LDA $ & 0.429  & 0.822  & 1.01  & 1.66   \\ \hline
  $NP-SIC-LDA $ & 0.214  & 0.410  & 0.505 & 0.83   \\ \hline
  $RPA-SIC-LSD$ & 1.23   & 2.13  & 2.62 & 4.15   \\ \hline
  $GB-SIC-LSD $ & 0.994  & 1.77  & 2.18 & 3.50   \\ \hline
  $NP-SIC-LSD $ & 0.836  & 1.39  & 1.70 & 2.67  \\ \hline
  $VMC$~\cite{Buendia06}  &0.619 & 1.27  & 1.43& 2.03 \\ \hline
  Clementi and           &0.842 & 1.74  & 2.26 & 4.04 \\ 
  Hoffmann~\cite{Clementi95}  & &  & & \\ \hline
\end{tabular}
\caption{Correlation energies (Har) from DFT calculations using various 
  correlation functionals and their self-interaction corrected 
  versions for closed-shell atoms.}
\label{tab:DFT2}
\end{table}
\end{center}

In addition to power series forms, it is also possible to represent
logarithmic terms in a pair density context.  A simple functional form for
the correlation energy per particle common to the approximations of Gell-Mann
and Brueckner~\cite{GellMann57} (GB), Nozi\`eres and Pines~\cite{Nozieres58}
(NP), and the random-phase approximation~\cite{Mahan00} (RPA) may be written
as
\begin{equation}
  e_c(r_s) = A_c + B_c \mbox{ln} r_s,
\label{eqn:e_c_dft}
\end{equation}
or taking advantage of the definition of $r_s$,
\begin{equation}
 \frac{4 \pi r_s^3}{3} = \frac{1}{\rho}
\end{equation}
\begin{equation}
e_c(\rho) = A + B \mbox{ln} \rho.
\label{eqn:ec_dft}
\end{equation}
The first term (constant $A$) converts analogously to Eqs.  (\ref{eqn:T_ser})
and (\ref{eqn:t_g}).  In the local-density approximation for DFT without SIC
the constant term would correspond to $A N$, where $N$ denotes the number of
particles.  The second term, however, can be written in a PDFT form as
\begin{equation}
  B \mbox{ln} \rho = C + \frac{D}{2N}\int d{\bf r}  d{\bf r'} \tilde{n}_x({\bf r},{\bf r'})\mbox{ln}|{\bf r}-{\bf r'}|,
\end{equation}
and hence one can arrive at a PDFT analog of the form
\begin{equation}
e_c[\tilde{n}_x] = \tilde{A} + \frac{\tilde{B}}{2N}\int d{\bf r}  d{\bf r'}
\tilde{n}_x({\bf r},{\bf r'})\mbox{ln}|{\bf r}-{\bf r'}|.
\label{eqn:ec_pdft}
\end{equation}

The derivation of the constants proceeds analogously to the derivation in
Eqs. (\ref{eqn:T_ser}) and (\ref{eqn:t_g}) (one uses the HF pair density, and
the definition of the Fermi wave-vector).  The values of the constants for GB
and NP for both cases (DFT and PDFT) are given in Table \ref{tab:constants}.
The values of the constant $\tilde{A}$ and $\tilde{B}$ are fixed by requiring
that the correlation energies be equal to the known values for the
homogeneous case (i.e. Eq. (\ref{eqn:ec_pdft}) reproduces Eq.
(\ref{eqn:ec_dft})).  The derivation of the relation between the constants is
given in the Appendix.

The use of the SIC exchange-correlation hole in Eq.  (\ref{eqn:ec_pdft}) has
two important consequences.  It has been pointed
out~\cite{Perdew81b,Gorling93} that under homogeneous coordinate scaling the
logarithmic term in the approximate correlation energy functionals such as
GB, NP, and RPA, diverges as the scaling parameter $\lambda$ is made
infinite.  This divergence is avoided in Eq. (\ref{eqn:ec_pdft}) due to the
use of the exchange hole.  Indeed, coordinate scaling shifts the functional
in Eq. (\ref{eqn:ec_pdft}) by an amount proportional to $\int d{\bf r} d{\bf
  r'} \tilde{n}_x({\bf r},{\bf r'})\mbox{ln}\lambda$, which integrates to
zero in the case of an orthonormal basis set.  This is advantageous since it
has also been shown that the actual correlation functional scales to a
bounded constant~\cite{Levy89,Perdew81b,Gorling92}.  For two-electron atoms
the constant has been calculated: -0.0467Har~\cite{Ivanov99,Perdew96}.  In our
case this constant is zero for one class of our constructed functionals (Eq.
(\ref{eqn:ec_pdft2})) and it corresponds to the constant $\tilde{A}$ shown in
Table \ref{tab:constants} for the other (Eq.  (\ref{eqn:ec_pdft3})).  The
divergence would arise had we not applied the SIC to the exchange hole.

Another consequence is that self-interaction is avoided by construction.  The
first term in Eq.  (\ref{eqn:ec_pdft}) is therefore zero, and the constant
term $\tilde{A}$ does not contribute.  This can be shown by considering Eqs.
(\ref{eqn:f}) and (\ref{eqn:f_HF}).  When the exponent $\Gamma$ in Eq.
(\ref{eqn:f_HF}) is taken to be zero, and the exponent $a$ in Eq.
(\ref{eqn:f}) is taken to be one it follows that $b=0$.  In this case a
constant term in the correlation functional again corresponds to an
integration over the exchange hole (Eq.  (\ref{eqn:x_hole})) giving a
contribution of zero.  The disappearance of the constant term is also present
in the SIC of Lundin and Eriksson~\cite{Lundin01}.  In the standard DFT
version of the correlation functional the constant term also does not
contribute if the SIC procedure of Perdew and Zunger~\cite{Perdew81} is
applied to the functional {\it within the local density approximation} (for
an example of an application of the SIC in an LDA context see Ref.
~\onlinecite{Madjet01}), since the constant term in the correlation energy of
the electron gas (which is defined as energy per particle) corresponds to a
term consisting of an integral over the density to the first power.  Applying
the SIC to such an integral leads to cancellation.  The problem does not
arise when the SIC is formulated in the local-spin density (LSD)
approximation, since the constant is scaled~\cite{Misawa65} in the SIC term
which is subtracted from the uncorrected functional.  Below we also compare
using the two different versions of SIC for the constant term.

It should be emphasized that in both cases the cancellation of the constant
term is an artifact of the self-interaction procedure used, which, in this
case estimates the self-interaction component to be as large as the exchange
constant itself.  The constant in Eq.  (\ref{eqn:e_c_dft}), $A_c$, arises
from direct and exchange interactions~\cite{GellMann57}, and it would be a
severe approximation to discard it.  One possible way around this problem in
our formalism, in principle, is to relax the condition $a=1$.  Such a
procedure would, however, significantly worsen the scaling of the method,
since explicit evaluation of the two-body density would be necessitated.
Instead we apply the standard LSD based SIC procedure of Perdew and
Zunger~\cite{Perdew81} to $\tilde{A}$ in the correlation functionals, which
amounts to an additive term of $\tilde{A} N /2$.  Applying an LSD based SIC
procedure to the constant term is not inconsistent with the SIC procedure we
use, since our SIC corrected exchange hole $\tilde{n}_x$ is a sum over
different spin-components.

Another point to emphasize is that although the exchange hole includes
correlations between electrons with parallel spins only, the energy
functionals constructed according to our procedure account, at least in
principle, for all interactions that are included in the approximate
high-density correlation energy functionals (RPA, GB, NP).  The interactions
accounted for are determined by the functionals to which our PDFT based
functionals are made to correspond.  By analogous logic, in DFT the
correlation energy functional depends on the density only, a one-body object
without correlations, yet it accounts for correlation by virtue of being
required to reproduce the energetics of approximations to correlation in the
HEG.  

Our correlation energy functional (total as opposed to per particle) thus
looks like
\begin{eqnarray}
  E_c^I[\tilde{n}_x] = + \frac{\tilde{B}}{2}\int d{\bf r}  d{\bf r'}
  \tilde{n}_x({\bf r},{\bf r'}) \mbox{ln}|{\bf r}-{\bf r'}|.
\label{eqn:ec_pdft2}
\end{eqnarray}
When the constant term $\tilde{A}$ is subject to LSD based SIC treatment the
correlation energy functional becomes
\begin{eqnarray}
  E_c^{II}[\tilde{n}_x] = \frac{\tilde{A} N}{2}+ \frac{\tilde{B}}{2}\int 
d{\bf r}  d{\bf r'}
\tilde{n}_x({\bf r},{\bf r'})
\mbox{ln}|{\bf r}-{\bf r'}|.
\label{eqn:ec_pdft3}
\end{eqnarray}

\section{Calculation details}

\label{sec:calculation}

While the above formalism amounts to a modified HF, here we perform a normal
HF calculation and calculate the correlation energies using the HF orbitals.
Thus our calculations amount to a perturbative treatment.  This treatment was
also suggested by Levy~\cite{Levy87}.  Given that for our test cases (closed
sub-shell atoms) the correlation energies are ca. 1\% of the total energy,
the perturbative approach can be expected to be a good approximation.

We have performed atomic HF calculations using the most recent
version~\cite{Gaigalas96} of the atomic program written by
Fischer~\cite{Fischer86}.  Subsequently we have used the orbitals obtained
from the atomic program to calculate the correlation energy contributions
given by the PDFT analogs that we constructed.  For comparison we have also
calculated the correlation energies from the DFT analogs of GB and NP.  The
GB~\cite{GellMann57} and NP~\cite{Nozieres58} correlation functionals are
based on a diagrammatic perturbation summation for the HEG.  While the GB is
an expansion valid in the limit of high-density, the NP functional is an
interpolation formula approximately valid for all density
ranges~\cite{Mahan00,Dreizler90}.

For reference purposes we have calculated the correlation energies within
DFT.  In Tables \ref{tab:DFT} and \ref{tab:DFT2} we present the correlation
energies for a sequence of closed-shell and closed sub-shell atoms calculated
using the GB, NP, and RPA forms.  The results of Clementi and
Hoffmann~\cite{Clementi95} and the variational Monte Carlo results of
Buend\'ia {\it et al.}~\cite{Buendia06} are also shown.  As expected the
correlation energies are overestimated due to the absence of SIC.
Application of SIC based on LDA leads to an underestimation of the
correlation energy, whereas the LSD based SIC leads to an improvement over
both previous methods.  When improved correlation energy functionals are used
the correlation energy is quantitatively recovered~\cite{Perdew81}.

\section{Results and analysis}

\label{sec:results}

\begin{center}
\begin{table}
\begin{tabular}{|l|l|l|l|l|l|}
  \hline
  $E_c$    & He    &   Be   &  Ne   &   Mg  & Ar    \\ \hline
  $RPA-I$    & 0.0   & 0.00133 & 0.0704 & 0.0755 & 0.156 \\ \hline
  $GB-I$    & 0.0   & 0.00133 & 0.0704 & 0.0755 & 0.156 \\ \hline
  $NP-I$    & 0.0   & 0.000661 & 0.0351 & 0.0376 & 0.0776 \\ \hline
  $RPA-II $    & 0.0838   & 0.169 & 0.489 & 0.578 & 0.910 \\ \hline
  $GB-II $       & 0.0598   & 0.121 & 0.369 & 0.434 & 0.694 \\ \hline
  $NP-II $       & 0.0639   & 0.128 & 0.354 & 0.421 & 0.652 \\ \hline
  $VMC$~\cite{Buendia06}  & - & 0.073  & 0.346 & 0.372 & 0.576 \\ \hline
  Clementi and   &0.045 & 0.094  & 0.3870& 0.438 & 0.722 \\ 
  Hoffmann~\cite{Clementi95}  & &   & &  & \\ \hline 
\end{tabular}
\caption{Correlation energies (Har) from PDFT calculations using various 
  exchange correlation functionals.  For the meaning of
  the different functionals (I, II) see the text.  }
\label{tab:PDFT_I}
\end{table}
\end{center}

\begin{center}
\begin{table}
\begin{tabular}{|l|l|l|l|l|}
  \hline
  $E_c$    & Ca    &   Zn   &  Kr   &   Xe     \\ \hline
  $RPA-I$    & 0.166   & 0.362 & 0.340 & 0.829  \\ \hline
  $GB-I$    & 0.166   & 0.362 & 0.340 & 0.829  \\ \hline
  $NP-I$    & 0.0830   & 0.181 & 0.170 & 0.368  \\ \hline
  $RPA-II $    & 0.99   & 1.62 & 1.84 & 3.09  \\ \hline
  $GB-II $       & 0.763   & 1.26 & 1.42 & 2.44  \\ \hline
  $NP-II $       & 0.722   & 1.14 & 1.32 & 2.13  \\ \hline
  $VMC$~\cite{Buendia06}  &0.619 & 1.27  & 1.43& 2.03 \\ \hline
  Clementi and           &0.842 & 1.74  & 2.26 & 4.04 \\ 
  Hoffmann~\cite{Clementi95}  & &  & & \\ \hline
\end{tabular}
\caption{Correlation energies (Har) from PDFT calculations using various 
  exchange correlation functionals.  For the meaning of
  the different functionals (I, II) see the text.  }
\label{tab:PDFT_II}
\end{table}
\end{center}

In Table \ref{tab:PDFT_I} the correlation energies based on the PDFT analogs
of the RPA, GB, and NP correlation functionals are presented for a sequence
of closed sub-shell atoms He, Be, Ne, Mg, Ar (I refers to the functional
given in Eq.  (\ref{eqn:ec_pdft2})).  The calculated correlation energies
capture the qualitative trend for all three functionals.  An apparent
deficiency is the fact that the correlation energy for the helium atom is
zero.  This deficiency stems from the use of the SIC exchange-correlation
hole as the functional to construct the correlation energy functional.  The
SIC exchange correlation hole (Eq. (\ref{eqn:xc_hole})) for a non-interacting
system consists of interactions between electrons with parallel spins.  Since
the two electrons in a helium atom (ground state) are an anti-parallel pair
it can be expected that a correlation energy functional of the exchange
correlation hole give zero as a result.

We note that there is a qualitative similarity and at the same time a
quantitative difference between this short-coming of functional I (Eq.
(\ref{eqn:ec_pdft2})) and the spurious self-interaction in the case of the
original LDA density functionals.  In LDA it is a property of the {\it input
  function} (the one-body density) that gives rise to the spurious
contribution.  Since the one-body density is some finite positive-definite
function even for a single electron a functional that depends on it gives a
finite contribution in general.  The similar artifact of the SIC exchange
hole as an input function is that it lacks terms with anti-parallel
interactions, therefore it is not expected to give contributions when a
system with one single interaction between electrons of anti-parallel spins
is considered, such as helium.

In Table \ref{tab:PDFT_I} results for the approximately corrected version
(Eq. (\ref{eqn:ec_pdft3})) of the PDFT correlation energy are also presented
(RPA-II, GB-II, and NP-II).  All three functionals are improved considerably
compared to their values without the SIC Table \ref{tab:PDFT_I} and
quantitative (near exact) agreement is observed for the GB-II and NP-II
functionals when compared to the results of Clementi and
Hoffmann~\cite{Clementi95}.

The pair density ansatz GB-II appears to give the best results for the
sequence of atoms when compared to the results of Clementi and
Hoffmann~\cite{Clementi95}.  The results for GB-II are also consistently
better than the corresponding DFT results in Table \ref{tab:DFT}
(GB-SIC-LSD).  

The recent results of Buend\'ia {\it et al.}~\cite{Buendia06} are also
presented for comparison.  This study is based on a variational Monte Carlo
calculation using a correlated basis function.  Our results compare well with
the known results, and overall the agreement appears to be as good as the VMC
calculation of Buend\'ia {\it et al.}~\cite{Buendia06}.  The fact that in
our case the agreement worsens for small atom systems can be attributed to the
fact that the correlation functional is approximated from the HEG, a system
in the thermodynamic limit.

In Table \ref{tab:PDFT_II} results for a heavier sequence of closed sub-shell
atoms Ca, Zn, Kr, Xe are also presented.  The correlation energies of
Clementi and Hoffmann~\cite{Clementi95} are also shown for comparison.  While
our results are no longer in as good agreement for the correlation energy as
for the lighter atoms in Table \ref{tab:PDFT_I}, they appear to be close to
the results of the VMC calculation of Buend\'ia {\it et
  al.}~\cite{Buendia06}.  The results of Clementi and
Hoffmann~\cite{Clementi95} appear to be better reproduced by DFT in this case
(Table \ref{tab:DFT2}).  A potential source of error in our case is the fact
that we are using functionals based on the high-density limit of the
HEG~\cite{Mahan00,GellMann57,Nozieres58} which may not be a good
approximation for electrons farther from the nucleus.

\section{Conclusion}

\label{sec:conclusion}

We developed an electronic structure method based on pair density functional
theory that is simple to implement.  We have shown that exact energy
functionals can in principle be constructed in terms of the Hartree-Fock pair
density, and that the representability question can be circumvented.  Our
formalism is closely related to the work of Levy on Hartree-Fock density
matrix functional theory~\cite{Levy87}.  In applications approximations have
to be developed to account for correlation.  In our method correlation is
accounted for by auxiliary pair-potentials between electrons in a
Hartree-Fock setting, thus the method scales also as Hartree-Fock.  The
correlation approximations tested give quantitative agreement for atomic test
cases with other methods.  In DFT there are many possible options for
correlation functionals which have been developed over the past few decades,
often with specific chemical or physical situations in mind.  The close
agreement for the two simple PDFT correlation functionals developed and
tested in this work indicate that there is potential in developing
correlation functionals for PDFT as well.  Since the pair density is a
quantity depending on the coordinates of two particles it can be expected to
give a better overall, more robust description of correlation than the ones
used in one-body DFT.

\appendix

\begin{center}
{\bf ACKNOWLEDGMENTS}
\end{center}

\label{sec:acknowledgments}

We acknowledge using the resources of CINECA.  BH would like to thank Filippo
De Angelis, Stefano Baroni, Roberto Car, Sandro Scandolo, Gaetano Senatore,
and Giacinto Scoles for helpful discussions.

\begin{center}
{\bf APPENDIX}
\end{center}

\label{sec:appendix}

In this appendix we show how to determine the relation between the constants
for a correlation energy for the HEG of the
form~\cite{Mahan00,GellMann57,Nozieres58}
\begin{equation}
e_c(\rho) = A + B \mbox{ln} \rho.
\label{eqn:eqna1}
\end{equation}
The two-body functional form we assume for the correlation functional can be
written as 
\begin{equation}
e_c(\tilde{n}_x) = \tilde{A} + \frac{\tilde{B}}{2N} \int d {\bf r}{\bf
  r'}
   \tilde{n}_x({\bf r},{\bf r'}) \mbox{ln} |{\bf r}-{\bf r'}|.
\label{eqn:eqna2}
\end{equation}
Using the fact that 
\begin{equation}
\tilde{n}_x(|{\bf r}-{\bf r'}|) = \rho^2  [g(|{\bf r}-{\bf r'}|)-1],
\label{eqn:eqna3}
\end{equation}
we can rewrite Eq. (\ref{eqn:eqna2}) as 
\begin{equation}
\epsilon(\tilde{n}_x) = \tilde{A} + 2 \pi \tilde{B}\rho \int r^2 dr
   \tilde{n}_x(r) \mbox{ln} |r|.
\label{eqn:eqna4}
\end{equation}
Scaling the coordinate $r$ as
\begin{eqnarray}
x &=& k_f r \\
k_f &=& (3 \pi^2 \rho)^{\frac{1}{3}} \nonumber
\label{eqn:eqna5}
\end{eqnarray}
results in 
\begin{eqnarray}
e_c(\tilde{n}_x) &=& \tilde{A} + \frac{2 \tilde{B}}{3 \pi} \int x^2 dx
   [g(x)-1] \mbox{ln} |x| \\
&& + \frac{\tilde{B}}{6} \mbox{ln} 3 \pi^2 \rho. \nonumber
\label{eqn:eqna6}
\end{eqnarray}
Using the form of $g(x)$~\cite{Mahan00}, the relation between the constants
can then be shown to be
\begin{eqnarray}
\tilde{A} &=& A - \frac{4B}{\pi}\int x^2 dx
   [g(x)-1] \mbox{ln} |x| \\
&& - B \mbox{ln} 3 \pi^2 \nonumber\\
\tilde{B} &=& 6B. \nonumber
\label{eqn:eqna7}
\end{eqnarray}



\begin{thebibliography}{10}

\bibitem{Hohenberg64} P. Hohenberg and W. Kohn, {\it Phys. Rev.} {\bf
    136} B864 (1964).

\bibitem{Kohn65} W. Kohn and L.~J. Sham, {\it Phys. Rev.} {\bf
    140} B1133 (1965).

\bibitem{Parr89} R.~G. Parr and W. Yang, {\it Density-Functional Theory of
    Atoms and Molecules} Oxford University Press (1989).

\bibitem{Dreizler90} R.~M. Dreizler and E.~K.~U. Gross, {\it Density
    Functional Theory: an Approach to the Quantum Many-Body Problem} Springer
  (1990).

\bibitem{Szabo96} A. Szabo and N. Ostlund, {\it Modern Quantum Chemistry} 1st Ed., Dover (1996).

\bibitem{Young01} D.~C. Young, {\it Computational Chemistry: A Practical
    Guide to Real-World Problems} John Wiley \& Sons (2001).

\bibitem{Senatore94} G. Senatore and N.~H. March, {\it Rev. Mod. Phys.} 
{\bf 66} 445 (1994).

\bibitem{Ceperley80} D.~M. Ceperley and B.~J. Alder, {\it
    Phys. Rev. Lett.} {\bf 45} 566 (1980).

\bibitem{Mahan00} G.~D. Mahan, {\it Many-Particle Physics} 3rd Ed., Kluwer
  Academic (2000).

\bibitem{Mattuck67} R.~D. Mattuck, {\it A Guide to Feynman Diagrams in the
    Many-Body Problem}, Dover (1992).

\bibitem{March67} N.~H. March, W.~H. Young, and S. Sampanthar, {\it The
    Many-Body Problem in Quantum Mechanics}, Dover (1995).

\bibitem{GellMann57} M. Gell-Mann and K.~A. Brueckner {\it Phys. Rev.} {\bf
    106} 364 (1957).

\bibitem{Nozieres58} P. Nozi\`eres and D. Pines {\it Phys. Rev.} {\bf
    111} 442 (1958).

\bibitem{Perdew81} J.~P. Perdew and A. Zunger {\it Phys. Rev. B} {\bf 23}
  5048 (1981).

\bibitem{Foresman96} J.~B. Foresman and \AE. Frisch {\it Exploring Chemistry
    with Electronic Structure Methods} Gaussian, Inc. (1995-1996).

\bibitem{Payne79} P.~W. Payne{\it J. Chem. Phys.} {\bf 71}490 (1979).

\bibitem{Levy79} M. Levy {\it Proc. Acad. Nat. Sci. USA} {\bf 76} 6062
  (1979).

\bibitem{Ziesche96} P. Ziesche {\it Int. J. Quant. Chem.} {\bf 60} 149 (1996).

\bibitem{Levy01} M. Levy and P. Ziesche {\it J. Chem. Phys.} {\bf 115} 9110
  (2001).

\bibitem{Furche04} F. Furche {\it Phys. Rev. A} {\bf 70} 022514
  (2004).

\bibitem{Hetenyi04} B. Het\'enyi, L. Brualla, and S. Fantoni {\it Phys. Rev.
    Lett.} {\bf 93} 170202 (2004).

\bibitem{Nagy02} A. Nagy {\it Phys. Rev. A} {\bf 66} 022505 (2002).

\bibitem{Nagy04} A. Nagy and C. Amovilli {\it J. Chem. Phys.}
  {\bf 121} 6640 (2004).

\bibitem{Nagy06} A. Nagy {\it Int. J. Quant. Chem.} {\bf 106} 1043 (2006).

\bibitem{Percus05} J.~K. Percus {\it J. Chem. Phys.}  {\bf 122} 234103 (2005).

\bibitem{Ayers05} P.~W. Ayers and M. Levy {\it Chem. Phys. Lett.} {\bf
    415} 211 (2005).

\bibitem{Ayers06a} P.~W. Ayers and E.~R. Davidson {\it Int. J. Quantum Chem.}
  {\bf 106} 1487 (2006).

\bibitem{Ayers06b} P.~W. Ayers, S. Golden, and M. Levy {\it J. Chem. Phys.}
  {\bf 124} 054101 (2006).

\bibitem{Ayers07} P.~W. Ayers and S. Liu {\it Phys. Rev. A} {\bf
    75} 022514 (2007).

\bibitem{Gonis96} A. Gonis, T.~C. Schultess, J. van Ek, and P.~E.~A. Turchi {\it Phys. Rev. Lett.} {\bf 77} 2981 (1996).

\bibitem{Gonis97} A. Gonis, T.~C. Schultess, P.~E.~A. Turchi, and J. van Ek {\it Phys. Rev. B} {\bf 56} 9335 (1997).

\bibitem{Pistol04} M.~E. Pistol {\it Chem. Phys. Lett.} {\bf 400} 548 (2004).

\bibitem{Higuchi07} M. Higuchi and K. Higuchi {\it Physica B} {\bf 387} 117 (2007).

\bibitem{Zumbach85} G. Zumbach and K. Maschke, {\it
    J. Chem. Phys. } {\bf 82} 5604 (1985).

\bibitem{Levy87} M. Levy in {\it Density Matrices and Density Functionals},
  Ed. R. Erdahl and V.~H. Smith Jr., D. Reidel, (1987).

\bibitem{Mazziotti06} D.~A. Mazziotti, {\it
    Acc. Chem. Res.} {\bf 39} 207 (2006).

\bibitem{Coleman00} A.~J. Coleman and V.~I. Yukalov, {\it Reduced Density
    Matrices: Coulson's Challenge} Springer-Verlag, New York (2000).

\bibitem{Coleman63} A.~J. Coleman, {\it Rev. Mod. Phys.} {\bf 35} 668 (1963).

\bibitem{Gorling93} A. G\"orling and M. Levy, {\it
    Phys. Rev. B} {\bf 47} 13105 (1993).

\bibitem{Perdew81b} J. P. Perdew, E. R. McMullan, and A. Zunger, {\it
    Phys. Rev. A} {\bf 23} 2785 (1981).

\bibitem{Gorling92} A. G\"orling and M. Levy, {\it
    Phys. Rev. A} {\bf 43} 4637 (1992).

\bibitem{Levy89} M. Levy, {\it Int. J. Quant. Chem.} {\bf S23} 617 (1989).

\bibitem{Kimball76} J. Kimball {\it Phys. Rev.} {\bf
    14} 2371 (1976).

\bibitem{Krotscheck71} E. Krotscheck and M.~L. Ristig, {\it Phys. Lett. } {\bf 48A} 17 (1971).

\bibitem{Fantoni74} S. Fantoni and S. Rosati, {\it Nuovo Cimento} {\bf 10} 545
  (1974).

\bibitem{Foulkes01} W.~M.~C. Foulkes, L. Mitas, R.~J. Needs, and G.
  Rajagopal, {\it Rev. Mod. Phys.} {\bf 73} 33 (2001).

\bibitem{Lundin01} U. Lundin and O. Eriksson, {\it
    Int. J. Quant. Chem.} {\bf 81} 247 (2001).

\bibitem{Bobel83} G. B\"obel and G. Cortona, {\it
    J. Phys. B: At. Mol. Opt. Phys.} {\bf 16} 349 (1983).

\bibitem{Rae75} Rae A.~I.~M. {\it Mol. Phys.} {\bf 29} 467 (1975).

\bibitem{Ivanov99} S. Ivanov, K. Burke, and M. Levy, {\it J. Chem. Phys.}
  {\bf 110} 10262 (1999).

\bibitem{Perdew96} J.~P. Perdew, K. Burke, and M. Ernzerhof, {\it Phys. Rev.
    Lett. } {\bf 77} 3865 (1996).

\bibitem{Madjet01} M. E. Madjet, H. S. Chakraborty, and J.-M. Rost, {\it
    J. Phys. B: At. Mol. Opt. Phys.} {\bf 34} L345 (2001).

\bibitem{Misawa65} S. Misawa, {\it Phys. Rev.} {\bf 140} A1645 (1965).

\bibitem{Gaigalas96} G. Gaigalas and C.~F. Fischer, {\it
    Comp. Phys. Comm.} {\bf 98} 255 (1996).

\bibitem{Fischer86} C.~F. Fischer, {\it
    Comp. Phys. Rep.} {\bf 3} 273 (1986).

\bibitem{Clementi95} E. Clementi and D.~W.~M. Hoffmann, {\it
    J. Mol Struct.-Theochem} {\bf 330} 17 (1995).

\bibitem{Buendia06} E. Buend\'ia, G\'alvez, and A. Sarsa, {\it
    Chem. Phys. Lett.} {\bf 428} 241 (2006).


\end{thebibliography}
\end{document}